\documentclass{WileyMSP-template}

\usepackage{amsmath}

\begin{document}

\pagestyle{fancy}
\rhead{\vspace{0.5cm}}
 
\title{Unbounded Systematic Error in Thin Film Conductivity Measurements}

\maketitle


\author{Yongyi Gao},
\author{Hio-Ieng Un},
\author{Yuxuan Huang},
\author{Henning Sirringhaus},
\author{Ian E. Jacobs*}

\begin{affiliations}
	Cavendish Laboratory, University of Cambridge
    J J Thomson Avenue, Cambridge, CB3 0US\\
	Email Address: ij255@cam.ac.uk

\end{affiliations}



\begin{abstract}

Electrical conductivity is the most fundamental charge transport parameter, and measurements of conductivity are a basic part of materials characterization for nearly all conducting materials. In thin films, conductivity is often measured in four bar architectures in which the current source and voltage measurement are spatially separated to eliminate systematic error due to contact resistance. Despite the apparent simplicity of these measurements, we demonstrate here that the four bar architecture is subject to significant systematic error arising from the finite conductivity of the metal electrodes. Remarkably, these systematic errors can in some cases become \textit{unbounded}, producing arbitrarily high measured conductivity at modest true film conductivities, within the range relevant to emerging thin film thermoelectric materials such as conducting polymers. These unbounded errors, which can occur even in properly conducted four-point measurements of patterned films, likely explain literature reports of extremely high conductivities in conducting polymers, and can lead to anomalous scaling in temperature dependent studies, potentially leading to incorrect interpretation of the relevant charge transport mechanism. We characterize the device geometric factors that control these errors, which stand partially at odds with those required for accurate Seebeck coefficient measurements. Our analyses allow us to identify reliable device architectures that provide small systematic errors for conductivity and Seebeck coefficient while still assuring a low measurement resistance, critical to reducing noise in thermal voltage measurements. These findings provide important guidelines for accurate measurements in the growing field of thin-film thermoelectric materials. 

\end{abstract}


\section{Introduction}
The first law of thermodynamics dictates that all energy conversion processes generate heat. An attractive approach to reduce global energy use is to utilize this waste heat, for instance by conversion to electrical energy. Thermoelectric generators (TEGs)\textemdash solid state devices which convert thermal gradients into electrical power\textemdash are an increasingly promising technology for waste heat harvesting applications.\cite{zevalkink_practical_2018,beretta_thermoelectrics_2019,shi_advanced_2020} Conjugated polymer based thermoelectrics in particular have seen significant research interest over the past decade owing to their potential application in flexible and wearable applications.\cite{russ_organic_2016,wang_flexible_2019} These materials are now among the highest-performing room temperature thermoelectrics.\cite{wang_multi-heterojunctioned_2024}

The performance of a thermoelectric material is commonly expressed by the dimensionless figure of merit
\[
ZT = \frac{S^{2}\sigma T}{\kappa},
\]
where $S$ denotes the Seebeck coefficient, $\sigma$ the electrical conductivity, $T$ the absolute temperature, and $\kappa$ the thermal conductivity. Improving $ZT$ therefore requires either a reduction of $\kappa$ or an increase of the power factor $PF = S^{2}\sigma$; consequently, precise and reproducible measurements of $\kappa$, $S$, and $\sigma$ are essential for reliable material evaluation and device optimisation.\cite{zevalkink_practical_2018,chetty_best_2024,bernhard_dorling_towards_2026} Thermal conductivity measurements are notoriously difficult and prone to systematic error.\cite{toberer_advances_2012} For thin-film polymer thermoelectrics, it is usually the in-plane component of $\kappa$ that is required for characterization of ZT. These measurements require careful design of both the device and measurement protocol to remove the contribution from the substrate, which is often thicker and more thermally conductive than the material of interest.\cite{cahill_thermal_1990,dames_measuring_2013,wang_brief_2019} Similarly, Seebeck measurements are known to be prone to errors arising from challenges in precisely measuring the experimentally realizable temperature difference across the device, which is typically only a few Kelvin, as well as the resulting thermal voltage, which is generally less than 1 mV.\cite{martin_high_2010,wang_contributed_2018} Due to the difficulty in both $\kappa$ and $S$ measurements, there has been considerable effort within the organic thermoelectrics community to identify systematic errors in these measurements, and to propose best practices for future work.\cite{bernhard_dorling_towards_2026}

In comparison, electrical conductivity\textemdash the last measurement required to calculate $ZT$\textemdash is relatively straightforward. Although a variety of measurement architectures are in widespread use within the community, including four bar, Hall bar, linear four-point probe, and van der Pauw, in most cases there is little discrepancy between measurements conducted on different architectures.\cite{noauthor_resistivity_2005} Systematic errors in these measurements are generally understood to be a result of geometric errors, for instance, by stray current in unpatterned devices, or from contact resistances in two-point measurement architectures. The errors introduced by these sources are typically small (less than a factor of 2).\cite{duong_measuring_2024} Nonetheless, there have been a number of reports of very high conductivity in doped conjugated polymers, exceeding the values of many metals, that remain challenging to understand.\cite{vijayakumar_bringing_2019,dhoot_beyond_2006,durand_single_2022,zeng_optimizing_2022} Ref \cite{vijayakumar_bringing_2019} in particular, which reports electrical conductivity \textgreater $10^5$ S/cm in aligned films of the polymer PBTTT (Poly[2,5-bis(3-tetradecylthiophen-2-yl)thieno[3,2-b]thiophene]), has inspired several followup studies. However, these studies themselves are not entirely consistent, with some reporting conductivities of 4000-5000,\cite{huang_design_2021,zhu_enhancing_2024} and others as low as $\sim$500 S/cm.\cite{tanaka_facile_2024} Spectroscopy and scattering studies have not shown dramatic differences between these samples; no clear mechanism has been proposed for the orders-of-magnitude or deviation between results from different groups.

We argue that both the high conductivities reported in Refs. \cite{vijayakumar_bringing_2019}-\cite{zeng_optimizing_2022} and the comparatively low conductivity reported in Ref. \cite{tanaka_facile_2024} likely arise from systematic errors intrinsic to four bar and Hall bar device geometries, which can surprisingly lead to unbounded systematic errors even in carefully performed four-point measurements of patterned films. These artifacts, which arise from incomplete shorting of the lateral potential across the device by the metal electrodes, to our knowledge have not been discussed previously in the literature of doped conjugated polymers or organic electrochemical transistors. Consequently, these artifacts appear to be widespread and are not limited to specific research groups,\cite{dhoot_beyond_2006,thomas_role_2018,vijayakumar_bringing_2019, tanaka_thermoelectric_2020,  ito_charge_2021,zeng_optimizing_2022,durand_single_2022,tanaka_facile_2024} manifesting most often as an \textit{underestimate} of true film conductivity or transconductance when using high aspect-ratio (FET-like) device geometries. In addition to complicating inter‑study comparisons, these systematic errors are inherently temperature-dependent, and can dominate the temperature dependence of conductivity, potentially leading incorrect conclusions regarding the charge transport mechanism. 

In the context of thermoelectrics characterization, optimizing device geometry to eliminate these systematic conductivity errors can also under some circumstances increase the systematic error of Seebeck measurements. Using a coupled thermal–electrical model\cite{antonova_finite_2005} that includes heat conduction, Seebeck and Peltier effects, Joule heating, and the Thomson relation, we quantify measurement errors associated with common electrode and probe configurations and systematically evaluate how device geometry and probe positioning bias the extracted Seebeck coefficient and conductivity. Our findings identify trade-offs between layouts that favor accurate measurement of $S$ versus $\sigma$, and provide practical guidelines to minimize systematic measurement error in the characterization of thin‑film conducting materials.

\section{Methods}
\subsection{Finite Element Method Simulations}
We model a system of two coupled partial differential equations describing the local temperature and potential. The electrical current density is
\begin{equation}
\mathbf{J} = -\sigma \!\left(\nabla V + S\,\nabla T\right)
\end{equation}

where $\sigma$ is the electrical conductivity, $S$ is the Seebeck coefficient, $t$ is the layer thickness, and $V$ and $T$ are the electric potential and temperature to be solved, respectively. The heat flux is
\begin{equation}
\mathbf{q} = -\kappa \,\nabla T + \Pi\,\mathbf{J}
\end{equation}

where $\kappa$ is the thermal conductivity and $\Pi = S T$ is the Peltier coefficient (Thomson relation). Joule heating provides the heat source through energy conservation under electric field $\mathbf{E}$:

\[
- \nabla\!\cdot\mathbf{q} \;=\; \mathbf{J}\!\cdot\!\mathbf{E}.
\]

Combining these relations with conservation of energy and charge yields the coupled thermo‑electric equations \cite{antonova_finite_2005,ebling_multiphysics_2009}:

\begin{equation}
	-\vec{\nabla} \cdot \left(\left(\kappa +\sigma S^{2}T\right)\vec{\nabla}T\right)
	-\vec{\nabla} \cdot\left(\sigma  S T\vec{\nabla}V\right)
	= \sigma \left(\left(\vec{\nabla}V\right)^2 + S\, \vec{\nabla}T \cdot \vec{\nabla}V\right)
	\label{eq:thermoelectric 1}
\end{equation}

\begin{equation}
	\vec{\nabla} \cdot  \left(\sigma S\, \vec{\nabla}T\right)
	+ \vec{\nabla} \cdot \left(\sigma \, \vec{\nabla}V\right) = \mathbf{J_{s}}
	\label{eq:thermoelectric 2}
\end{equation}

We solve the system of coupled PDEs (Equations \ref{eq:thermoelectric 1} and \ref{eq:thermoelectric 2}) in two dimensions using the finite-element method (FEM) as implemented in the MATLAB R2025b Partial Differential Equation Toolbox. In  two dimensions, the conductivities $\sigma$ and $\kappa$ become sheet conductivities, and are calculated by summing the product of conductivity and thickness for each layer present in the region, i.e. $\sigma = \sum_i \sigma_i t_i$ and $\kappa = \sum_i \kappa_i t_i$, where the layers $i$ correspond to gold, glass, and the thin film of interest (e.g. a doped conjugated polymer). The effective Seebeck coefficient of each region is weighted by the conductance of each layer present, $S = \sum_i S_i \sigma_i t_i / \sum_i  \sigma_i t_i $.  To model aligned films, we allow $\sigma$ and $\kappa$ to be anisotropic. The electrical conductivity of the metal electrodes are taken to be that of gold thin films,\cite{chopra_electrical_1963} $\sigma_E=1.25\times10^5$ S/cm, consistent with our experimental measurements. A full table of simulation parameters are given in Table S1 (see Supporting Information).

In conductivity simulations, the electrical current source is applied by setting $\mathbf{J_{s}} = \pm J_{meas}/A_{contact}$ within the source and drain probe contact regions with $A_{contact} = 100$ $\mu m^2$ the area of the probe contact region. Elsewhere $\mathbf{J_{s}}$ is uniformly 0. Insulating Neumann boundary conditions are assumed ($\nabla V = 0$) for the electrical PDE, while Dirichlet boundary conditions ($T=300$ K) are assumed for the thermal PDE. For Seebeck simulations, the simulation is modified such that $\mathbf{J_{s}} = 0$ everywhere. Dirichlet boundary conditions are applied to the left and right sides of the thermal PDE to establish a temperature gradient, with no heat flux allowed across the top and bottom edges ($\nabla T = 0$). Results are checked for mesh convergence, typically requiring a maximum triangle size equal to that of the smallest characteristic dimension in the model in the vicinity of small features; for safety margin we use a maximum triangle size 1/3 this dimension. We stress that in all simulations, the film is patterned, and that the errors identified here do not result from uncertainty in the dimensions of the conducting channel or current spreading outside the defined channel regions.

\subsection{Experimental Methods}
\subsubsection{Materials}
PBTTT-C12 (Mn=30 KDa, Mw=43.5 kDa, PDI=1.45) was synthesized as previously reported.\cite{huang_design_2021} BMP TFSI (1-Butyl-1-methylpyrrolidinium bis(trifluoromethanesulfonyl)imide, 99.9\% was purchased from Solvionic. Anhydrous FeCl3 (\textgreater99.99\% trace metals basis) was purchased from Sigma-Aldrich. Anyhdrous acetonitrile (AN) and 1,2-dichlorobenzene (DCB) were purchased from Romil (Hi-Dry, \textless20 ppm water).

\subsubsection{Sample preparation}
Samples fabricated as reported previously.\cite{huang_design_2021} Briefly, glass substrates (Corning Eagle XG, 700 $\mu$m thickness) were cleaned in an ultrasonic bath using a sequence of Decon 90 detergent solution, deionized water, acetone, and isopropanol, then dried under N$_2$ and exposed to oxygen plasma (300 W, 10 min). Cr/Au electrodes (5/25 nm) were then deposited by thermal evaporation through a shadow mask ($L_E = 220 \mu$m, $L_C = 580 \mu$m; see Figure \ref{fig:4PP EE distribution}a). PBTTT solutions (10 g/L DCB) were heated at 120$^{\circ}$C until fully dissolved, then cooled to 80$^{\circ}$C before use. Films were spin-coated at 1000 rpm, then annealed at 180$^{\circ}$C for 20 minutes and cooled slowly to room temperature. These films were then aligned on an RM-50 rubbing machine (E.H.C CO., LtD) at 155$^{\circ}$C using a microfiber cloth, then annealed again (190$^{\circ}$C 20 minutes) and cooled slowly to room temperature. The aligned films were then doped by ion exchange\cite{jacobs_high-efficiency_2022} (100 mM BMP TFSI / 1 mM FeCl$_3$ in AN, 60 seconds) then washed with excess AN. All sample preparation steps were performed under nitrogen (\textless 5 ppm O$_2$, \textless 10 ppm H$_2$O).

\subsubsection{Measurement}
Four point conductivity measurements were performed using an Agilent 4155B semiconductor parameter analyzer in probe configuration 4 (see Figure \ref{fig:4PP EE distribution}). Films were patterned using a toothpick, and the film width was measured by microscope images calibrated against the known device dimensions $L_E$ and $L_C$. All measurements, patterning, and imaging were performed under nitrogen atmosphere (\textless 5 ppm O$_2$, \textless 10 ppm H$_2$O). Average film thickness (60 nm) was measured by profilometery (Bruker DekTak XT) after melting the films to decrease roughness.

\subsection{Simulation of four bar device conductivity measurement}
\subsubsection{Effect of probe configuration on conductivity measurement}
\label{sec:probeconfigs}

\begin{figure}[b!]
	\centering
	\includegraphics[width=0.8\textwidth]{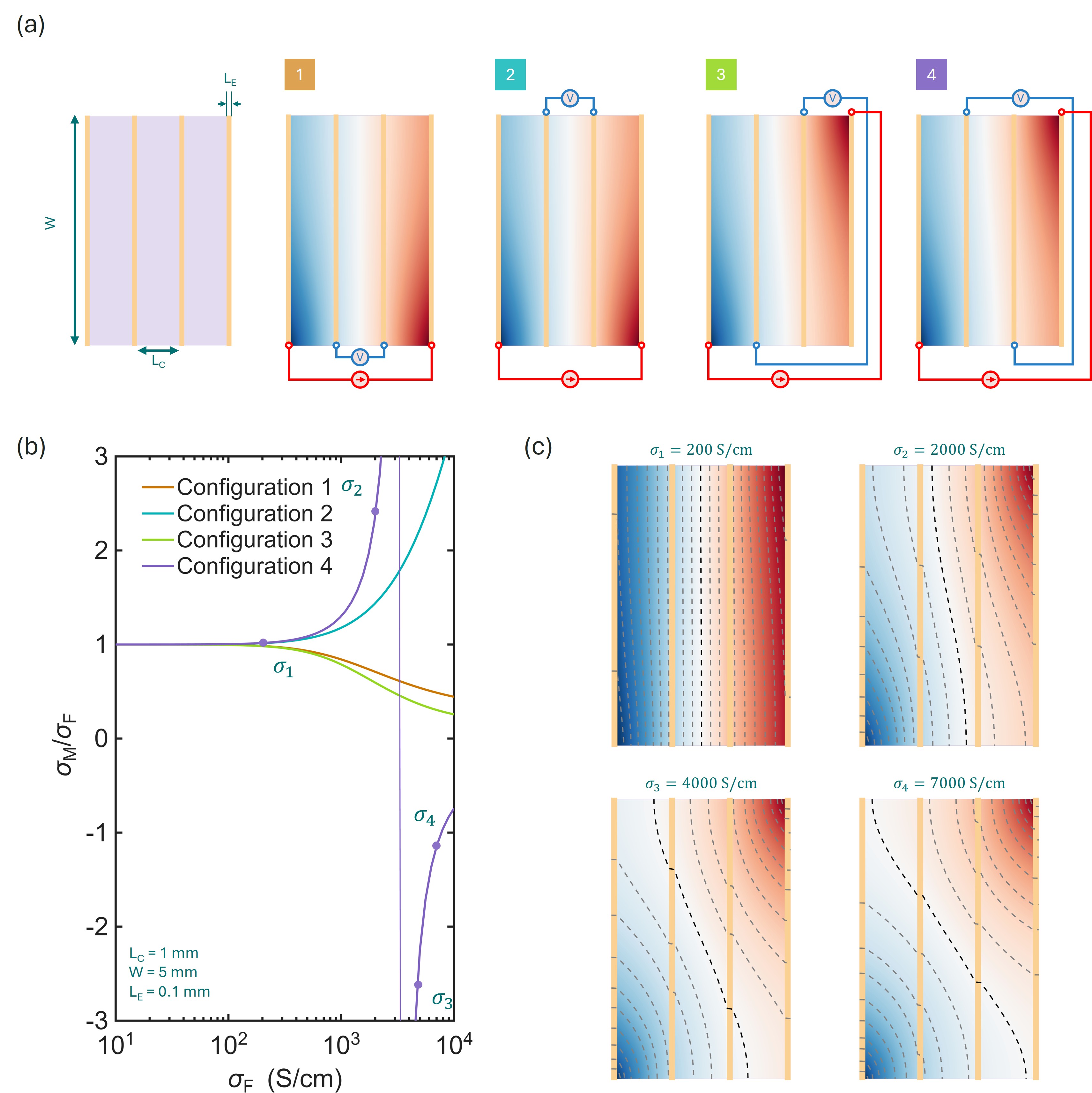}
	\caption{(a) Schematic of four‑point probe (FPP) conductivity measurement with thin‑film conductivity of 1000 S/cm. Current (red) is injected at the source and drain electrodes (red hollow circles). Voltage is measured at the middle voltage‑probe electrodes (blue hollow circles). Four configurations correspond to different choices of current and voltage probe positions. Color scale (blue-white-red) indicates the potential generated within the device by a fixed measurement current $J_{meas}$. (b) Ratio of measured conductivity to true conductivity ($\sigma_{M} / \sigma_{F}$) versus film conductivity for the four configurations. (c) Potential maps for configuration 4 at several film conductivities.}
	\label{fig:4PP EE distribution}
\end{figure}

Figure \ref{fig:4PP EE distribution} shows the results of our FEM conductivity model using a standard four bar device architecture, in which a fixed measurement current $J_{meas}$ is sourced between the outer electrode pair, and the potential difference $\Delta V$ is measured at the inner electrode pair. Ideally, the high conductivity of the metal electrodes should short the potential across the width of the device, resulting in a uniform current density flowing at all locations in the film. Under these conditions, the measured film conductivity $\sigma_M$ is calculated as
\begin{equation}
\sigma_M = \frac{L_{C}}{Wt_F} \frac{J_{meas}}{\Delta V}
\end{equation} 
where $L_{C}$ is the distance between each pair of electrodes, $W$ is the device width, and $t_F$ is the film thickness. 

Critically, in real experiments, electrical connections to the device are made via small pins or probe needles. We model these connections by sourcing the electrical current between two small regions (10 $\mu m$ square) on the outer electrode pair, and measuring $\Delta V$ between two discrete locations near the ends of the voltage probe electrodes. Figure \ref{fig:4PP EE distribution}a shows potential maps for four point configurations in a typical four bar structure ($L_C = 1$ mm, $L_E = 0.1$ mm, $W = 5$ mm), for a film conductivity of 1000 S/cm, typical of high-performance organic thermoelectric materials.\cite{wang_multi-heterojunctioned_2024,tang_solution-processed_2022,bubnova_semi-metallic_2014,jacobs_high-efficiency_2022} One might expect the probe locations should not appreciably affect the potential within the device, since the conductivity of the gold electrodes ($>10^5$ S/cm) still 2 orders of magnitude greater than the film conductivity. Nonetheless, we clearly observe that the potential maps depend on probe placement, and in all cases show considerable gradients across the device width, equivalent to a non-uniform current density throughout the device. These gradients lead to changes in $\Delta V$, and therefore produce systematic errors in the measured conductivity.

Figure \ref{fig:4PP EE distribution}b shows the relative systematic error\textemdash the ratio of the measured conductivity $\sigma_M$ to true conductivity $\sigma_F$\textemdash as a function of the true film conductivity for each probe configuration shown in Figure \ref{fig:4PP EE distribution}a. At low conductivity, the device behaves ideally and all probe configurations provide the same, accurate, measured conductivity. However, above about 200 S/cm, we begin to observe appreciable error. The onset of non-ideality corresponds to the conductivity at which the resistance between source and drain electrodes, $R_{SD} = \frac{3}{\sigma_{F}} \frac{L_{C}}{W t_{F}}$ approaches the resistance across the width of the device, primarily defined by the electrode resistance $R_{E} = \frac{1}{\sigma_{E}} \frac{W}{L_E t_{E}}$ (dimensions indicated in Figure \ref{fig:4PP EE distribution}a). Ideal device operation requires $R_E / R_{SD} \ll 1$, corresponding to
\begin{equation}
\boxed{
\frac{\sigma_F}{\sigma_E} \frac{W^2 t_F}{L_E L_C t_E} \ll 1
\label{eq:limit}
}
\end{equation}
Equation \ref{eq:limit} can be used to quickly calculate the safe operating regime for four bar devices; for instance, in the geometry modelled in Figure \ref{fig:4PP EE distribution}, a film conductivity of 200 S/cm corresponds to a value of $\sim 0.25$. This analysis also provides an explanation for the appearance of non-ideality at conductivities much lower than that of the metal electrodes. For typical four bar devices, the device aspect ratio $W/L_C \approx 1-10$. However, the relevant aspect ratio for $R_E$, which is measured across the width of the device, is $L_E/W$. For very thin electrodes with short $L_E$ this  ratio can be very small; for instance, it is 1/50 in the modelled device structure. Together, these geometric factors, which act to reduce $R_{SD}$ and increase $R_E$, lead to $R_E = R_{SD}$ at approximately $\sigma_F = \sigma_E / 100$. The requirement that $R_E$ is small relative to $R_{SD}$ increases the onset of non-ideality to approximately $\sigma_F = \sigma_E / 500$ in the modelled device. We stress that while Equation \ref{eq:limit} is general, the conductivity at which the device deviates from ideality is highly sensitive to geometry, including layer thicknesses.

As seen in Figure \ref{fig:4PP EE distribution}b, in the non-ideal operating regime some configurations produce systematic underestimates of the true conductivity, while others produce systematic overestimates. We can understand these behaviors via the potential maps in Figure \ref{fig:4PP EE distribution}a. In probe configurations 1 and 2, there is a competition between parallel conduction pathways flowing directly from source to drain, and pathways across the device further from the source and drain probes. These latter pathways will have higher resistance due to the additional series contribution of $R_E$, resulting in a higher current density flowing on the side of the device nearest the source and drain probes. On this edge of the device, the potential profile is actually identical to the ideal case where $R_E = 0$, because the path directly from source to drain does not travel along the width of the device. Therefore, in the absence of contact resistance, configuration 1 will produce the voltage probe readings expected from the device geometry (i.e. 1/3 and 2/3 of the source-drain voltage). However, because the contribution of $R_E$ forces more current to flow on the side of the device closest to the source-drain probes, for fixed $J_{meas}$ the source-drain voltage must increase relative to an ideal device. In configuration 1, this in turn increases $\Delta V$, resulting in a systematic underestimate of the true conductivity. In configuration 2, the voltage is probed on the opposite side of the channel, where the current density is lower relative to the ideal case. This results in a lower $\Delta V$, producing a systematic overestimate of the true conductivity.

In configurations 3 and 4, current is applied at opposite ends of the source and drain, producing a potential gradient running from the lower left corner to the upper right corner of the devices. This off-axis potential gradient causes the voltage probes in configuration 3 to see a larger $\Delta V$, leading to an underestimate of the true conductivity. The opposite is true in configuration 4\textemdash the potential gradient rotation instead reduces $\Delta V$ and produces an overestimate. However, as can be seen in Figure \ref{fig:4PP EE distribution}b this configuration is more complex than the others and produces a \textit{negative} measured conductivity at high true film conductivities. 

Figure \ref{fig:4PP EE distribution}c shows the evolution of the device potential in configuration 4 for several film conductivities. The black dashed line shows the potential isocontour corresponding to a voltage halfway between the source and drain. At low conductivity, this isocontour runs straight down the middle of the device, indicating that all locations within the left voltage probe electrode are at a potential closer to the source (leftmost electrode), and similarly that the right voltage probe is at a potential closer to that of the drain (rightmost electrode). The resulting $\Delta V$ has the same sign as $J_{meas}$, producing a positive $\sigma_M$ as expected. As conductivity increases, the potential gradient rotates to lie along the line from source to drain probe locations, and this isocontour cuts diagonally across the device. Note that at high conductivities, the top edges of both voltage probe electrodes now sit on the same side of the isocontour, indicating that they both are at potential closer to the drain contact at top-left, while the bottom edges see a potential closer to the source contact at bottom right. In probe configuration 4, then, the left voltage probe reads a potential closer to the right-most electrode, and vice-versa, producing a negative $\Delta V$ and hence a negative measured conductivity. \textit{Critically, this implies that at some intermediate conductivity, the potential gradient must rotate such that a potential isocontour connects the voltage probe locations, corresponding to a $\Delta V$ of 0 and hence an infinite measured conductivity.}

Stated another way, even in perfectly uniform, patterned films, four bar devices can surprisingly produce unbounded overestimates of the true film conductivity. Because the error is a \textit{systematic} error, repeated measurements of the same or similar samples will give repeatable results. Remarkably, these errors can arise under quite benign conditions; in the device structure modeled, the measured conductivity diverges at $\sigma_F < 3000$ S/cm, within the conductivity range of many high performance polymeric conductors;\cite{bubnova_semi-metallic_2014, huang_design_2021, ke_highly_2023} doubling the device width moves this divergence to only $\sigma_F \approx 500$ S/cm. 

Counter-intuitively, it is the use of four point measurements\textemdash which are intended to reduce systematic error by removing contact resistance effects\textemdash that is responsible for the unbounded error. In two point measurements, there is no way to produce a $\Delta V$ of 0 when $J_{meas}$ is nonzero; it is the spatial separation of the voltage probes from the electrodes generating the potential gradient that allows for this possibility in four point measurements. Instead, in two point measurements, finite electrode resistances $R_E$ always increase the total resistance from source to drain relative to the ideal case ($R_E = 0$) and therefore always produce a systematic underestimate of the true conductivity, regardless of the probe configuration.

We are not aware of any prior discussion of this unbounded error in the scientific literature. Four bar devices with geometry similar to that modeled here, using probe configuration 4, appear to have been used in many of the reports of anomalously high conductivity in doped conjugated polymers.\cite{vijayakumar_bringing_2019,durand_single_2022,zeng_optimizing_2022} We believe that these high reported conductivities are likely the result of researchers inadvertently approaching the regime where the measured conductivity diverges. 

\subsubsection{Effect of device geometry}
\label{sec:geometry}

\begin{figure}[t!]
	\centering
	\includegraphics[width=0.8\textwidth]{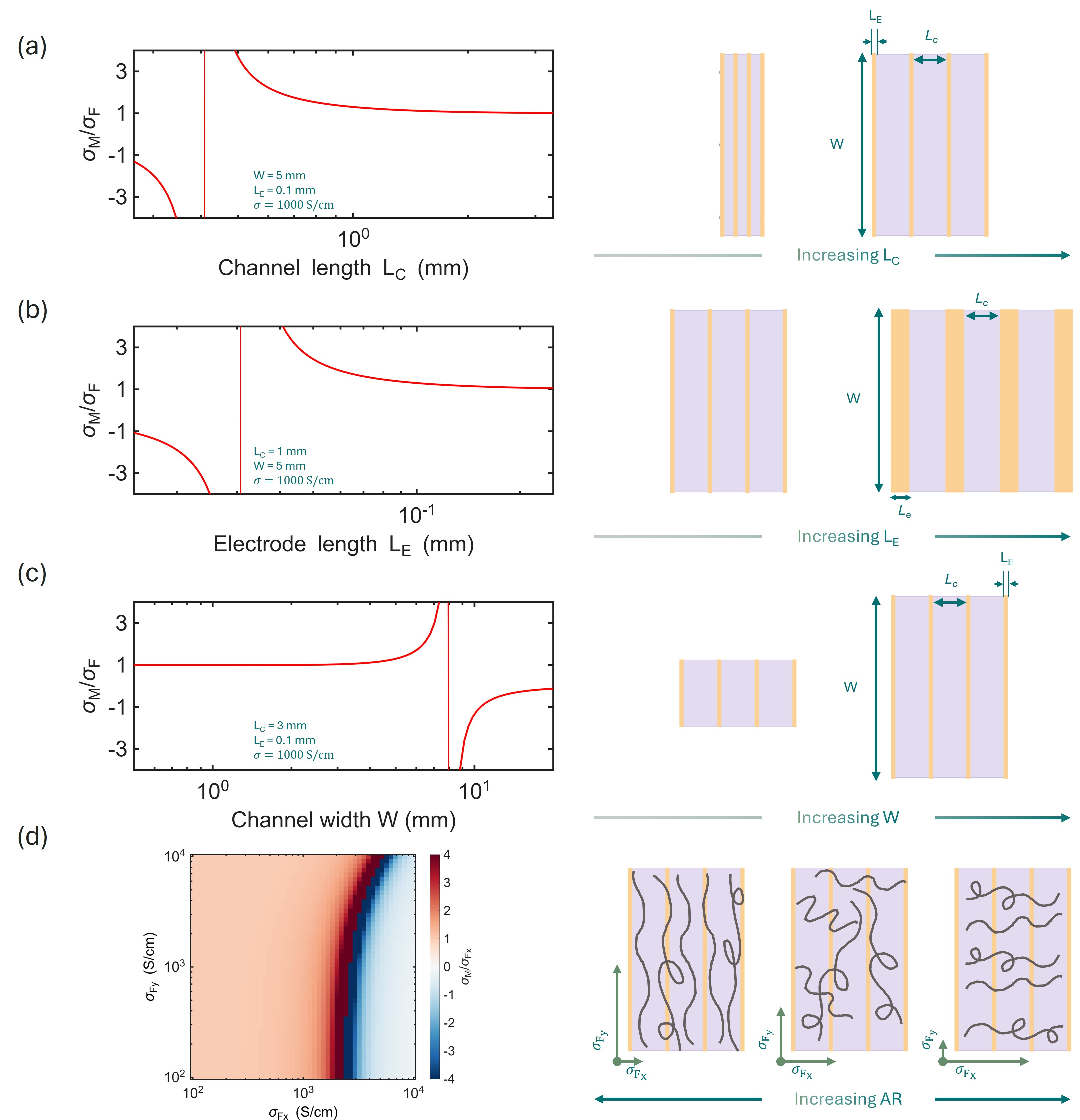}
	\caption{Geometric factors affecting systematic conductivity error $\sigma_M/\sigma_F$. a) Effect of varying channel length $L_C$, b) varying electrode length $L_E$, and c) varying channel width $W$. Non-varying geometric factors are identical to those in Figure \ref{fig:4PP EE distribution}. d) Map of systematic error for anistropic film conductivity; $\sigma_{Fx}$ is oriented along the channel length while $\sigma_{Fy}$ is oriented along the width. All simulations use configuration 4. Other parameters are as in Table S1 unless specified.}
	\label{fig:rc re ar effect}
\end{figure}

Thus far we have only discussed the effect of varying $\sigma_F$ on systematic error. However, varying any of the parameters in Equation \ref{eq:limit} that leads to a violation of the inequality will result in an error in the measured conductivity. To highlight the role of geometric factors, in Figure \ref{fig:rc re ar effect}a-c we plot the systematic measurement error in probe configuration 4 as the channel length $L_C$, electrode length $L_E$, and device width $W$ are varied. In each case all other parameters match those of the device in Figure \ref{fig:4PP EE distribution}; $\sigma_F$ is fixed at 1000 S/cm in all plots.

As shown in Figure \ref{fig:rc re ar effect}a, at long channel lengths $L_C$ the device behaves ideally because a large $L_C$ increases $R_{SD}$ but has no effect on $R_E$. Conversely, for short channels, we observe a divergence in $\sigma_M$ due to the reduction in $R_{SD}$. Scaling the electrode length $L_E$ (Figure \ref{fig:rc re ar effect}b) produces the same effect: long $L_E$ gives lower error, while short $L_E$ leads to conductivity divergence. In this case, reducing $L_E$ produces an increase in $R_E$, increasing the error, while leaving $R_{SD}$ unaffected. The error is particularly sensitive to the device width (Figure \ref{fig:rc re ar effect}c), as can be seen by the $W^2$ term in Equation \ref{eq:limit}. This quadratic dependence arises because increasing $W$ simultaneously reduces $R_{SD}$ and increases $R_E$. These results suggest that devices with very high-aspect ratio channels (large $W/L_C$),\cite{dhoot_beyond_2006,thomas_role_2018,ito_charge_2021} similar to those commonly used in field-effect transistor devices, are very likely to lead to non-ideal behavior when used to measure conducting films (e.g. doped conjugated polymers) unless the device is carefully designed to reduce $R_E$. To minimize conductivity measurement error devices should be long and narrow, corresponding to aspect ratios $W/L_C < 1$, and use relatively fat electrodes with large $L_E$.

\subsubsection{Effect of anisotropic conductivity}
Many organic semiconductors show anisotropic conduction due to polymer chain alignment or crystalline texture \cite{memon_alignment_2022}. In particular, several groups have reported that films of the polymer PBTTT can be aligned by solution shearing\cite{tanaka_facile_2024,zhu_enhancing_2024} or high-temperature rubbing,\cite{vijayakumar_bringing_2019,huang_design_2021} leading to anisotropy ratios $>10$. Upon doping, these aligned films were reported to achieve conductivities as high as $2\times 10^5$ S/cm \cite{vijayakumar_bringing_2019} in a device architecture similar to that modeled in Figure 1. In similar samples fabricated into Hall bar devices which we calculate should remain ideal at $>10^{6}$ S/cm, conductivities of only 4300 S/cm were reported.\cite{huang_design_2021} 

Figure \ref{fig:rc re ar effect}d shows the relative systematic error $\sigma_M/\sigma_F$ as the diagonal components of the film conductivity tensor, $\sigma_{Fx}$ and $\sigma_{Fy}$ are varied for the geometry shown in Figure \ref{fig:4PP EE distribution}. We observe that the error has little dependence on the perpendicular component, $\sigma_{Fy}$, except when $\sigma_{Fy}$ becomes very large. This is because, by symmetry, changes to $\sigma_{Fy}$ cannot not effect $R_{SD}$ but also have little effect on the resistance across the width of the device when $\sigma_{Fy}$ is small. An appreciable parallel contribution to $R_E$ by the film only occurs when $\sigma_{Fy}   \not \ll  \sigma_{E} \frac{L_E t_{E}}{L_C t_F}$. For the modelled geometry, $\sigma_{Fy}$ equals the right hand side at $\sim 2300$ S/cm, consistent with Figure \ref{fig:rc re ar effect}d, which shows the onset of dependence on $\sigma_{Fy}$ only above 500 - 1000 S/cm.

Conversely, scaling the component aligned along the length of the device, $\sigma_{Fx}$, produces an error very similar to scaling the isotropic conductivity (corresponding to a diagonal linecut of Figure \ref{fig:rc re ar effect}d). The argument is again similar\textemdash it is the $\sigma_{Fx}$ component of  $\sigma_{F}$ that is entirely responsible for changes to $R_{SD}$, while $\sigma_{Fy}$ has relatively little impact on the resistance across the device width except when $\sigma_{Fy}$ is very large. These results indicate that an enhancement in $\sigma_{Fx}$ by chain alignment to approximately 3000 S/cm should be sufficient to explain the reported conductivities of $2\times 10^5$ S/cm.

\subsubsection{Experimental validation of unbounded systematic conductivity error}
We validate the model described above by measuring the conductivity of a rub aligned PBTTT:TFSI film doped by ion exchange, following the procedure in Ref. \cite{huang_design_2021}. To vary the ideality of the device, we progressively reduce the device width by scratching off regions of the film between measurements, experimentally reproducing the curve shown in Figure \ref{fig:rc re ar effect}c. Device width is measured via microscope images taken between measurements.

\begin{figure}[t!]
	\centering
	\includegraphics[width=0.9\textwidth]{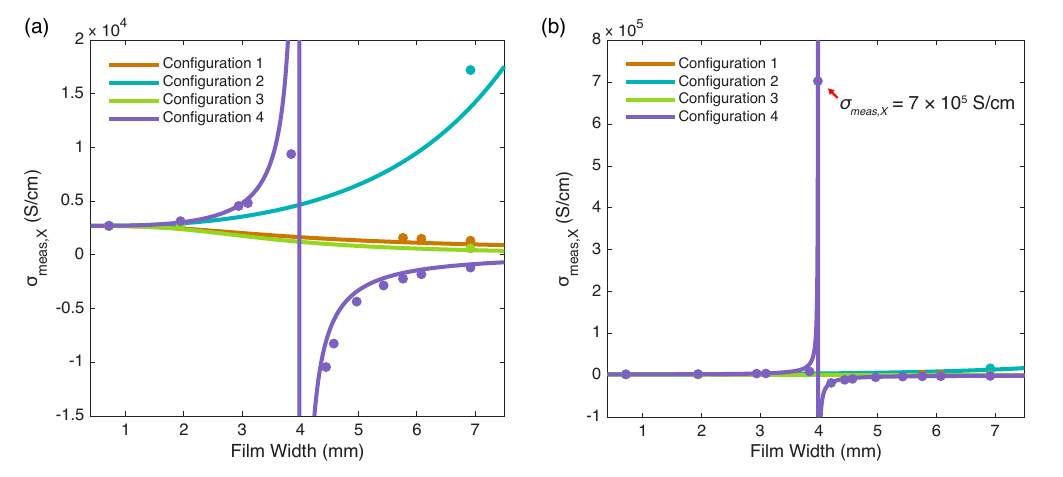}
	\caption{a) Experimental conductivity measurements of a highly doped rub-aligned PBTTT:TFSI film in four bar architecture ($L_E = 220 \mu$m, $L_C = 580 \mu$m). The conductivity is measured in different probe configurations as film width is progressively reduced by scratching off the film edges with a toothpick. Symbols are experimental data; solid lines are fit using our FEM model using film conductivities $\sigma_{Fx} = 2774$ S/cm, $\sigma_{Fy} = 440$ S/cm, and electrode conductivity $\sigma_{E} = 1.443 \times 10^5$ S/cm and experimentally measured film widths. b) The same data replotted on an expanded conductivity scale, showing an experimentally measured conductivity of $7\times10^5$ S/m near the divergence point.}
	\label{fig:exp}
\end{figure}

Figure \ref{fig:exp}a shows the measured conductivities for probe configurations 1-4 measured as a function of device width. Symbols correspond to experimental measurements, while solid lines indicate fits to the experimental data with $\sigma_{Fx} = 2774$ S/cm and $\sigma_{Fy} = 440$ S/cm, and $\sigma_{E} = 1.44\times 10^5$ S/cm; all other parameters are experimentally measured. These parameters are consistent with the values reported in Refs. \cite{huang_design_2021, chopra_electrical_1963}. Figure \ref{fig:exp}b shows the same plot on an expanded y-axis. Near the divergence point, we measure an apparent conductivity of $7\times10^5$ S/cm, corresponding to a systematic overestimate of the true conductivity by a factor \textgreater250. Therefore, the mechanism identified here does indeed appear to experimentally allow for unbounded systematic error.

\subsection{Temperature‑dependent conductivity measurement}
These errors can potentially also lead to deeper misinterpretations about the charge transport physics of these materials. Charge transport is often studied by examining the temperature dependence of the electrical conductivity. For instance, in metals at moderate temperatures, conduction is limited by phonon scattering, giving rise to a well-known $\sigma \propto 1/T$ relationship. In disordered semiconductors, charge transport is usually thermally activated, and models such as variable‑range hopping (VRH) are widely used. At temperature $T$, carriers hop between localized states, trading spatial distance against activation energy \cite{mott_electronic_1979}. Averaging the hopping probability yields the general Mott VRH form:

\begin{equation}
	\sigma(T) = \sigma_0 \exp\left[-\left(\frac{T_0}{T}\right)^{\frac{1}{d+1}}\right]
	\label{eq:VRH model}
\end{equation}

where $\sigma_0$ is a prefactor, $T_0$ a characteristic temperature, and $d$ the transport dimensionality. These relations enable fitting and physical interpretation of $\sigma(T)$ across temperature ranges. Such temperature dependent studies can give deep insights into the transport physics of these materials. However, in non-ideal devices, the magnitude of the systematic error will itself be temperature dependent, since $R_{SD}$ and $R_E$ both vary with temperature. These errors can therefore distort the scaling of conductivity with temperature. 

\begin{figure}
	\centering
	\includegraphics[width=0.95\textwidth]{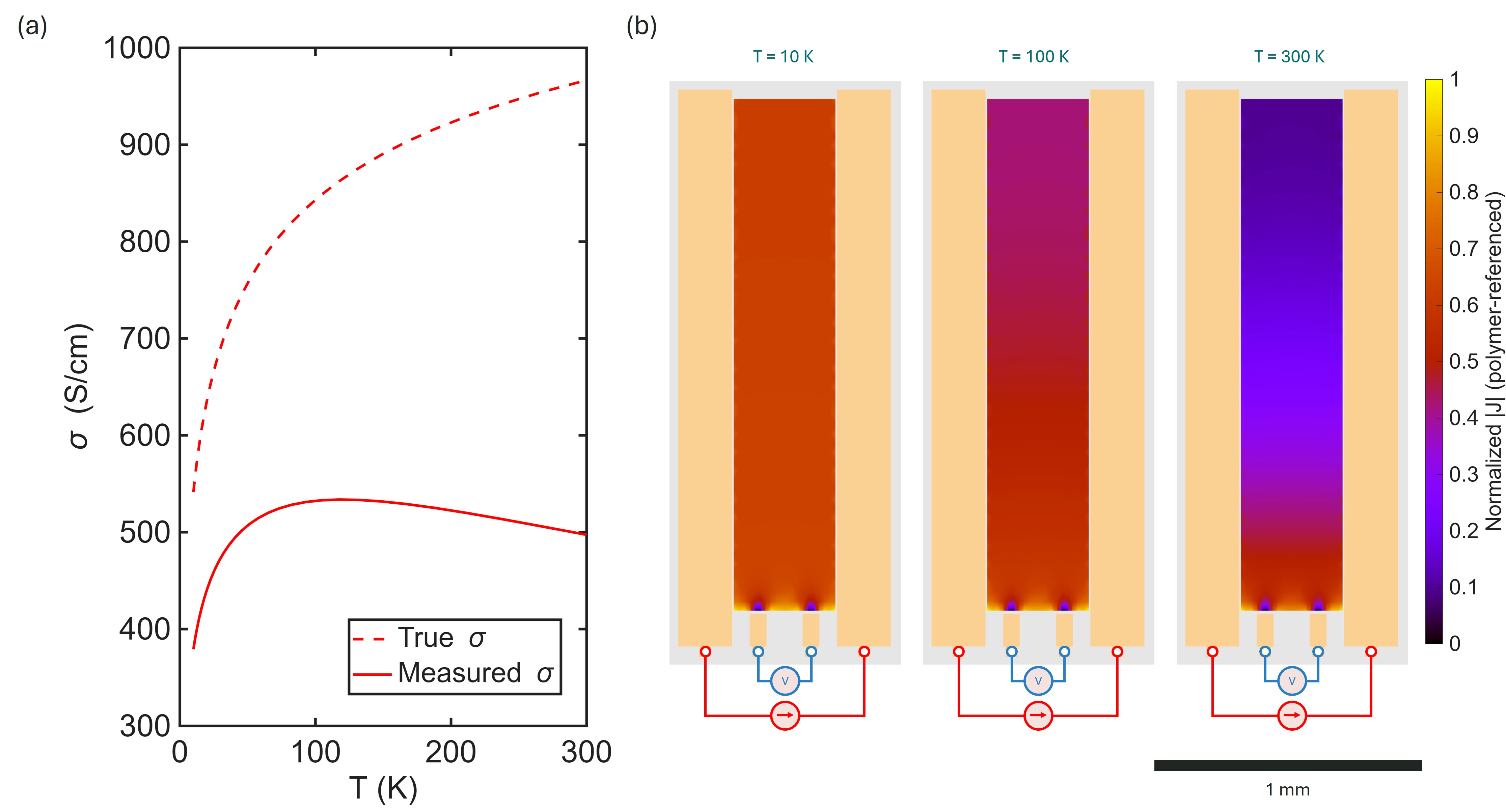}
	\caption{(a) Temperature‑dependent conductivity used as simulation input (3D Mott VRH). (b) Current density maps in the Hall bar geometry at 10 K, 100 K, and 300 K. Other parameters as in Figure \ref{fig:4PP EE distribution}. The gold electrode resistance also varies with temperature.}
	\label{fig:vrh model with T}
\end{figure}

To illustrate these geometry‑induced deviations in temperature‑dependent measurements, we model a Hall-bar like geometry, as commonly used in transport measurements,\cite{tanaka_thermoelectric_2020,tanaka_facile_2024} in configuration 1 (all probe contacts on one edge). We model the film conductivity assuming d = 3 (three‑dimensional Mott VRH) and compute \(\sigma_F(T)\) from Equation \eqref{eq:VRH model} (Figure \ref{fig:vrh model with T}). Electrode conductivity, which affects $R_E$, is modeled assuming phonon-limited metallic transport (Table S1). At 10K, the film conductivity is low, while the electrode conductivity is very high, producing a nearly uniform current density throughout the device (Figure \ref{fig:vrh model with T}b) and relatively small discrepancy between true and measured conductivities (Figure \ref{fig:vrh model with T}a). As the temperature is increased from 10 K to 300 K, the film conductivity rises while the electrode resistance simultaneously drops, reducing $R_{SD}$ and increasing $R_E$. As discussed above, this leads to a reduction in current density on the far side of the device, increasing $\Delta V$ and producing a progressively more underestimated conductivity as temperature increases. The resulting measured $\sigma_F(T)$, shown in Figure \ref{fig:vrh model with T}a, shows an increase in conductivity with decreasing temperature, appearing `band-like' despite the fact that the true transport mechanism in our model is localized VRH hopping. 

These errors are extremely difficult to identify experimentally because, as discussed earlier, in the probe configuration used here the voltage probe readings will appear ideal, reading the geometrically expected values relative to the source and drain voltages. There are therefore no obvious experimental signatures that the measurements are inaccurate. This subtle issue can then potentially lead to fundamentally incorrect interpretations of the underlying transport mechanism. Consideration of device geometry is therefore particularly important when performing temperature‑dependent transport measurements.

\subsection{Systematic errors in Seebeck measurement}

\begin{figure}[t]
	\centering
	\includegraphics[width=0.95\textwidth]{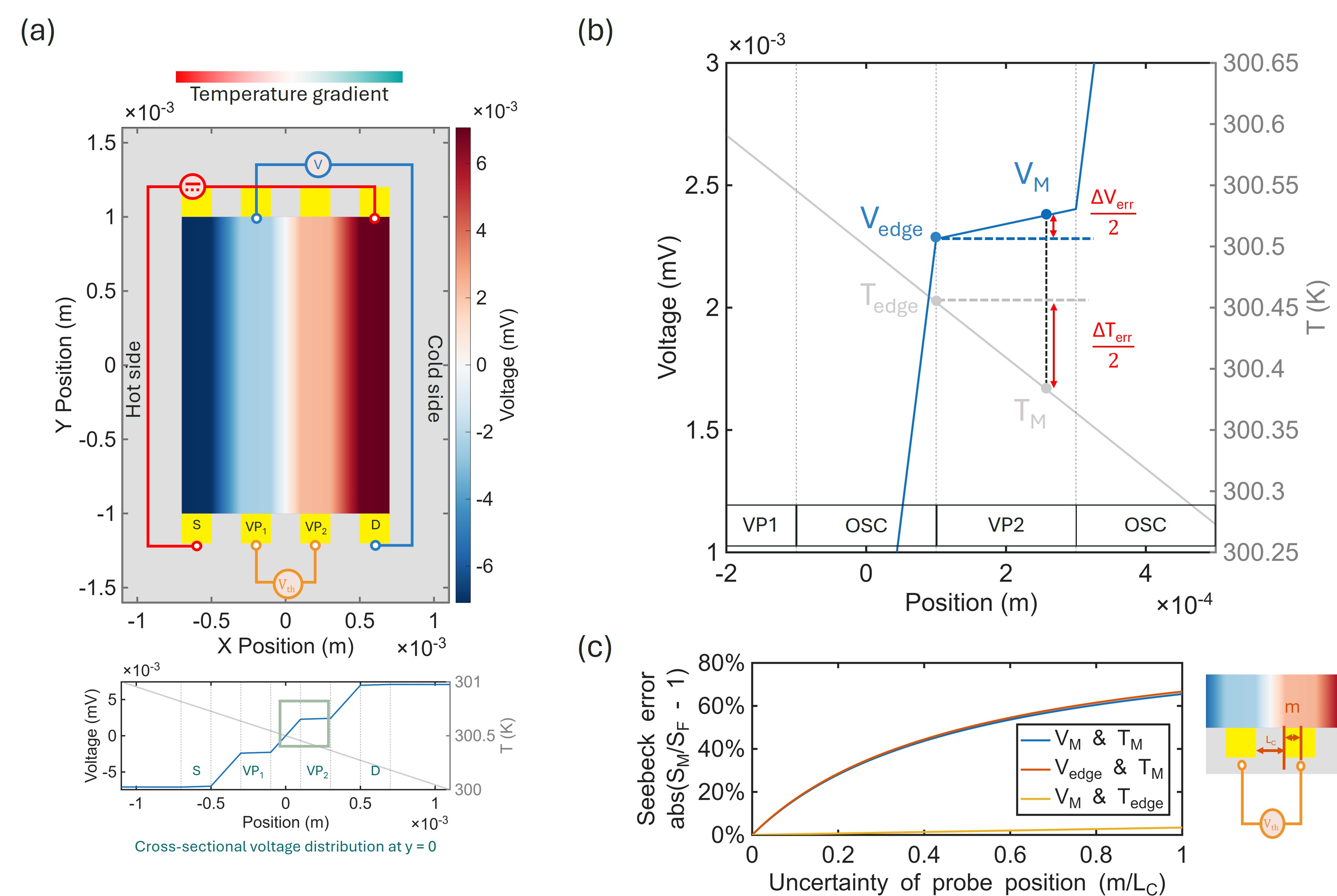}
	\caption{(a) Device geometry and dimensions. A temperature gradient is applied along $x$ (hot: 301 K; cold: 300 K). Thermal voltage is measured between voltage probes 1 and 2. The colormap shows the thermal‑voltage potential. (b) Voltage and temperature along $y=0$. (c) Enlarged view of (b). $V_{\text{measure}}$ and $T_{\text{measure}}$ denote probe‑measured values at the pad; $V_{\text{true}}$ and $T_{\text{true}}$ are values at the channel edge. $\Delta V$ and $\Delta T$ are the corresponding differences. (d) Absolute Seebeck error versus probe offset for different choices of $V$/$T$ in the calculation. Film $S=50\,\mu\mathrm{V\,K^{-1}}$ and $\sigma=500$ S/cm; other parameters as in Figure \ref{fig:4PP EE distribution}.}
	\label{fig:seebeck error source}
\end{figure}

Prior work by van Reenen et al. \cite{reenen_correcting_2014} showed that Seebeck errors can arise from current spreading through unpatterned films. In that work, it was shown that the ideal electrode geometry for Seebeck measurements consists of narrow strips, i.e. a device with large $W$ and short $L_E$. Unfortunately, as we have argued in Section \ref{sec:geometry}, such a geometry is likely to produce significant systematic error in conductivity measurements. 

To explore a possible solution to this dilemma, we first need to understand the error sources in Seebeck measurement. In our analysis, as throughout this paper, we assume films are patterned to exclude conduction pathways outside the device. In Figure \ref{fig:seebeck error source}, we simulate a four bar architecture using our steady-state 2D thermo–electric model at the device scale, without invoking a microscopic transport description. Our goal is to isolate the metrology contribution: how electrode geometry and probe placement bias the extracted Seebeck coefficient. 

Figure \ref{fig:seebeck error source}(a) shows the thermal voltage developed within the device when the simulation boundary conditions are set to produce a 1 K temperature difference. The resulting temperature gradient is almost perfectly linear (Figure \ref{fig:seebeck error source}a, bottom panel, right axis), since the substrate thermal conductance is large relative to that of the other layers. This thermal gradient produces a linear potential gradient within the channel regions of the device (Figure \ref{fig:seebeck error source}a, bottom panel, left axis). The film's Seebeck is modeled to be 50 $\mu$V/K, while the gold electrode Seebeck is set 1 $\mu$V/K. Calculating the Seebeck coefficient from the temperature and potential gradients in the channel closely matches the modeled $S$, confirming the internal consistency of our model.

It is generally considered best practice to measure the thermal voltages and temperatures at the same locations to ensure these measurements are self-consistent.\cite{bernhard_dorling_towards_2026} Nevertheless, even when both measurements are co-located, significant errors can still arise. These errors fall into two categories,  resulting from errors in the measured thermal voltage and measured temperature difference, respectively.  

When $T$ and $V$ measurements taken in the same locations, an accurate measurement of the film Seebeck coefficient requires co-located temperature and voltage pairs in which the film produces the only contribution to the measured thermal voltage. Experimentally, this implies that we should ideally measure $V$ and $T$ at the inner edges of voltage probe electrodes; we call these values $V_{edge}$ and $T_{edge}$, and the difference between the edge measurements at each electrode $\Delta V$ and $\Delta T$ such that $S_F = -\Delta V/ \Delta T$. In Figure \ref{fig:seebeck error source}b, we show a magnified plot of the simulated temperature and potential gradient around the interface between the channel and  the right voltage probe electrode, with the $V_{edge}$ and $T_{edge}$ labelled. However, in practice, the measurement probes can potentially be located in any position on top of the electrode, giving a slightly different pair of values $V_{M}$ and $T_{M}$. The difference between the measured and true values are the measurement error, $\Delta V_{err}$ and $\Delta T_{err}$. 

The origin of the error in $T_{M}$ is straightforward\textemdash when the probe position deviates from the ideal position at the inner edge of the electrode, we sample a different point $m$ along the temperature gradient. To estimate the average error, we assume the probes are placed at the middle of the electrodes, a distance $m = L_E / 2$ from the inner edge. Assuming a linear temperature gradient as modeled, the error in the measured temperature gradient is:

\begin{equation}
	\Delta T_{err} = \Delta T_{M} - \Delta T = \frac{2m}{L_{C}}\,\Delta T \implies \frac{\Delta T_{err}}{\Delta T} \approx \frac{L_E}{L_{C}}
	\label{eq:Terror}
\end{equation}

Note that Equation \ref{eq:Terror} only holds for linear temperature gradients, and is only approximate, due to the uncertainty in the probe position. The maximum error, which occurs when the probes are positioned at the outer edge of the electrodes, will be larger by a factor of 2. If the electrodes lie within a region where the temperature gradient is small, this error will be reduced.

The existence of a temperature gradient across the voltage probe contacts necessarily produces an additional thermal voltage, leading to a variation in the measured potential $V_{M}$ as the probe contact position is varied (Figure \ref{fig:seebeck error source}b. The magnitude of this additional thermal voltage can be calculated by first defining the effective Seebeck of the film-electrode bilayer:

\begin{equation}
	S_{\mathrm{eff}}= \frac{\sigma_{\mathrm{F}}t_{\mathrm{F}}S_{\mathrm{F}}+\sigma_{\mathrm{E}}t_{\mathrm{E}}S_{\mathrm{E}}}{\sigma_{\mathrm{F}}t_{\mathrm{F}}+\sigma_{\mathrm{E}}t_{\mathrm{E}}}
	\label{eq:effective S}
\end{equation}

where subscripts $F$ and $E$ indicate film and electrode, respectively, and $t$ is the thickness of each layer. The thermal voltage error is then $\Delta V_{err} = -S_{eff} \Delta T_{err}$, which when combined with Equation \ref{eq:Terror} and the definition of $S_{F}$ gives:

\begin{equation}
\frac{\Delta V_{err}}{\Delta V} = \frac{S_{eff}}{S_{F} } \frac{L_E}{L_{C}}
\label{eq:Verror}
\end{equation}

Comparing Equations \ref{eq:Terror} and \ref{eq:Verror}, the relative errors both depend on device geometry in the same way, but $\Delta V_{err} / \Delta V$ is scaled by $S_{eff} / S_F$ which is almost always $\ll 1$, as the electrode conductance is nearly always large relative to the film, while the electrode Seebeck is usually very small. Therefore, in Seebeck measurements where $V$ and $T$ are co-located, the dominant contribution to the Seebeck error is the temperature error:
\begin{equation}
	\frac{S_{meas}}{S_{F}}
	= \frac{1 + \dfrac{S_{eff}}{S_{F}}\dfrac{2m}{L_C}}
	{1 + \dfrac{2m}{L_C}}%
    \approx 
    \frac{1}{{1 + \dfrac{L_E}{L_C}}}
    \label{eq:seebeckErr}
\end{equation}

Figure \ref{fig:seebeck error source}c shows the calculated Seebeck error as a function of the probe position $m$ normalized to the channel length. When the Seebeck is calculated from $\Delta V_{M}$ and $\Delta T$, the error remains small regardless of probe placement, however when $\Delta T_{M}$ is used in the calculation we observe similar error regardless of whether $\Delta V_{M}$ or $\Delta V$ are used. In all cases, making the electrode length short relative to the channel length should lead to more accurate measurements.  However, our analysis suggests error can also be minimized by designing the measurement setup such that the temperature gradient drops primarily within the channel, with relatively constant temperature near the electrodes, or by independently characterizing the temperature gradient.

\subsection{Optimizing the geometry to minimize the measurement error}

\begin{figure}[t]
	\centering
	\includegraphics[width=1\textwidth]{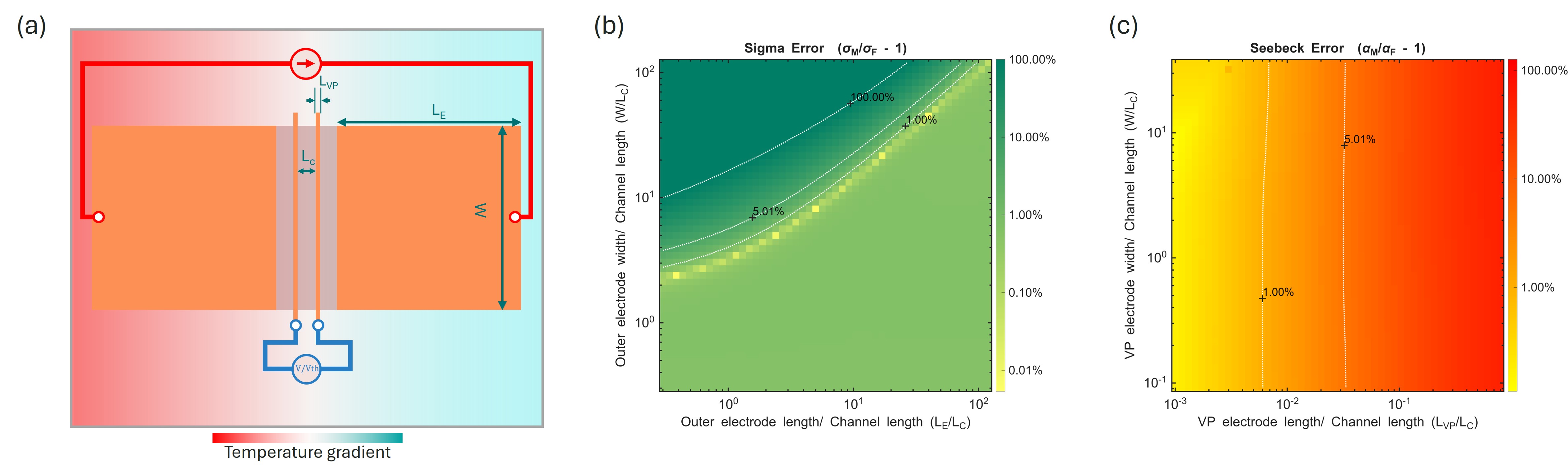}
	\caption{High conductivity (10000 S/cm) and low Seebeck coefficient (10 $\mu$V K$^{-1}$) simulation in four bar. (a) four bar layout illustration. The device has four characteristic dimensions; in our optimization we fix the channel length $L_C$ and vary the outer electrode length $L_E$, voltage probe electrode length $L_{VP}$, and device width $W$ relative to the channel length $L_C$. b) Four bar device conductivity measurement error as a function of outer electrode length $L_E$ and channel width $W$. Voltage probe electrodes are fixed at $L_{VP} = L_C/20$ c) Four bar device Seebeck measurement error versus electrode length and width. Outer electrodes are fixed at $L_E = 10L_C$}
	\label{fig:4pp and vdp}
\end{figure}

Our results suggest a conflict between the optimal architecture for conductivity measurements, which require long $L_E$, and the optimal architecture for Seebeck measurements, which require short $L_E$. We can solve this problem by recognizing that, for conductivity measurements, it is the $L_E$ of the outer electrode pair that is important\textemdash as long as these outer electrodes are able to eliminate any lateral potential gradients, the device will behave ideally, even if the inner electrode pair have very short $L_E$. Conversely, if the thermal voltage is measured at the inner electrode pair, the dimensions of the outer electrodes will not affect the Seebeck measurement. The optimal device will therefore have relatively wide outer electrodes and short inner electrodes as shown in Figure \ref{fig:4pp and vdp}. 

We simulate a film with high conductivity (10000 S/cm) and low Seebeck (10 $\mu$V K$^{-1}$). For both conductivity and Seebeck measurements the maximum error occurs in this regime: Conductivity error increases as $\sigma_F$ approaches $\sigma_E$, while the thermal voltage error increases when $S_F$ approaches $S_E$. The latter is however a weak effect, since as discussed above the dominant error is in the temperature difference, which to a good approximation does not depend on the properties of the film.  For $S$ measurements, $\Delta V$ and $\Delta T$ are taken at symmetric positions across the voltage pads (blue circles in Figure \ref{fig:4pp and vdp}a), and we record the worst-case error across the pad to quantify robustness. For $\sigma$ measurements, current is injected at symmetric positions on the center of the outer edges extended source/drain pads (red circles, Figure \ref{fig:4pp and vdp}a). 

To determine the outer electrode length $L_{E}$ and device width $W$, we fix $L_C = 1$ and plot the conductivity error as we vary these dimensions relative to $L_C$. Figure \ref{fig:4pp and vdp}b shows the results of these simulations. We observe that device widths up to $2L_C$ give low error acceptable even when $L_E < L_C$. Increasing the device width further requires an increase in $L_E$; for an aspect ratio $W/L_C=10$ we require $L_E = 10L_C$, corresponding to square outer electrodes as shown in Figure \ref{fig:4pp and vdp}a. High aspect ratio devices may be beneficial in in Seebeck measurements of resistive samples by reducing measurement noise which scales by approximately the square root of resistance.

To determine the inner electrode length $L_{VP}$, we perform a similar set of simulations varying $L_{VP}$ and $W$. As can be seen in Figure \ref{fig:4pp and vdp}, the Seebeck error has almost no dependence on $W$ and is primarily determined by $L_{VP}$, consistent with our analysis above. $L_{VP}$ can be reduced to improve the accuracy of the Seebeck measurements without penalty to the conductivity error, but will in practice be limited by the resolution of the fabrication process or the precision with which the electrical probes can be landed. We note that if the temperature gradient is linear across the full device and the temperature and voltage measurements are co-located, adding a pad to the end of the voltage probe electrodes will increase the Seebeck error, because it increases the uncertainty in the probe position and hence $\Delta T_{err}$. In practice we expect a typical architecture may be $L_E = L_C/20$, corresponding to a Seebeck error of about 5\%. 

In our simulations, we assume an equal distance $L_C$ between each pair of electrodes. However, in practice, having established optimized dimensions of $L_E$, $L_{VP}$, and $W$, it is usually beneficial to shift the voltage probe electrode pair outwards. This reduces both the systematic and random Seebeck error by decreasing $L_{VP}/L_C$ and increasing the temperature difference over channel, at the expense of potentially increasing thermal voltage measurement noise by increasing the channel resistance. The conductivity systematic error is not affected, as changing the position of the inner electrode pair does not affect $R_{SD}$, however the random error is decreased due to the increase in $\Delta V$. Other architectures, such as van der Pauw devices, can also achieve simultaneously low $S$ and $\sigma$ errors across a wide range of film conductivities, however the devices presented will likely be preferable for low-noise measurements due to their lower resistivity. See the Supporting Information for further discussion.

\section{Conclusion}
We have quantified geometry‑induced errors in four bar and Hall bar measurements of conductivity and Seebeck coefficient. It is found that non‑uniform potentials along thin‑film electrodes can unexpectedly produce large, and in some cases unbounded systematic errors when devices are wide or electrodes are narrow. These errors are intrinsically tied to the resistance of the film and electrodes and thus vary with temperature, distorting the true temperature dependence of the film conductivity. These effects, which to our knowledge were not previously considered, appear to be responsible for both over- and underestimates of electrical conductivity and transconductance in many literature reports. Identifying the issue in published data can be challenging due to a lack of detail regarding device structures. We encourage researchers to both validate their device architectures against Equation \ref{eq:limit} and to include a schematic and/or photograph of a representative device, including the location of probe contacts. We hope that these findings and framework for designing low-error devices for $\sigma$ and $S$ measurements will help to improve the reproducibility of results within the burgeoning thin film thermoelectrics community.

\medskip

\medskip
\textbf{Acknowledgements} \par 
I.E.J. gratefully acknowledges funding from a Royal Society University Research Fellowship (URF/R1/231287). H.S. is grateful for support
from a Royal Society Research Professorship (RP/R1/201082).

\medskip

%
\bibliographystyle{MSP}
\bibliography{references}

\end{document}